\documentclass[aps,prb,twocolumn,superscriptaddress,longbibliography]{revtex4-2}

\usepackage[T1]{fontenc}
\usepackage{amsmath,amssymb, bm}
\usepackage{graphicx}
\usepackage{siunitx}
\usepackage{hyperref}
\usepackage{xcolor}

\newcommand{\YIG}{yttrium iron garnet}
\newcommand{\GGG}{gadolinium gallium garnet}

\begin{document}

\title{Magnetoelastic coupling in stripe-domain states of yttrium iron garnet}

\author{Nimisha Arora}
\email{nimisha.arora@alumni.iitd.ac.in}
\affiliation{London Centre for Nanotechnology, University College London, London, WC1H 0AH, United Kingdom}

\author{Daniel Prestwood}
\affiliation{London Centre for Nanotechnology, University College London, London, WC1H 0AH, United Kingdom}
\affiliation{Department of Electronic and Electrical Engineering, University College London, London, WC1E 7JE, United Kingdom}
\affiliation{Blackett Laboratory, Department of Physics, Imperial College London, London SW7 2AZ, United Kingdom}

\author{Takashi Kikkawa}
\affiliation{Advanced Science Research Center, Japan Atomic Energy Agency, Tokai 319-1195, Japan}

\author{Eiji Saitoh}
\affiliation{Department of Applied Physics, The University of Tokyo, Tokyo 113-8656, Japan}
\affiliation{Institute for AI and Beyond, The University of Tokyo, Tokyo 113-8656, Japan}
\affiliation{RIKEN Center for Emergent Matter Science (CEMS), Wako 351–0198, Japan}
\affiliation{WPI-Advanced Institute for Materials Research, Tohoku University, Sendai 980-8577, Japan}

\author{Jack Gartside}
\affiliation{Blackett Laboratory, Department of Physics, Imperial College London, London SW7 2AZ, United Kingdom}
\affiliation{London Centre for Nanotechnology, Imperial College London, London, SW7 2AZ, United Kingdom}

\author{Will Branford}
\affiliation{Blackett Laboratory, Department of Physics, Imperial College London, London SW7 2AZ, United Kingdom}
\affiliation{London Centre for Nanotechnology, Imperial College London, London, SW7 2AZ, United Kingdom}

\author{Hidekazu Kurebayashi}
\email{h.kurebayashi@ucl.ac.uk}
\affiliation{London Centre for Nanotechnology, University College London, London, WC1H 0AH, United Kingdom}
\affiliation{Department of Electronic and Electrical Engineering, University College London, London, WC1E 7JE, United Kingdom}
\affiliation{WPI-Advanced Institute for Materials Research, Tohoku University, Sendai 980-8577, Japan}
\affiliation{Research Institute of Electrical Communication, Tohoku University, 2-1-1 Katahira, Sendai 980-8577, Japan}
\affiliation{Institute for Materials Research, Tohoku University, 2-1-1, Katahira, Sendai, 980-8577 Japan}
\date{\today}

\begin{abstract}
We study magnetoelastic coupling in stripe-domain magnetic states of \SI{3}{\micro\meter}-thick \YIG\ thin films grown on a \GGG\ substrate. Broadband ferromagnetic resonance reveals low-frequency stripe-domain magnon branches modulated by a field-independent phonon comb with a frequency spacing of $3.5\,\mathrm{MHz}$, matching the value predicted for confined thickness-shear modes of the GGG substrate. Analytical fitting yields coupling rates that vary between $0.33$--$0.54\,\mathrm{MHz}$ and cooperativities of order $10^{-1}$, indicating that the system is in the weak-coupling regime without resolvable avoided-crossing gaps. Magnon--phonon mode-overlap calculations using finite-element simulations show that the weak coupling arises from phase and domain-sign cancellation: the local magnetoelastic coupling is sizable, but more than $99\%$ of the coherent overlap cancels across the stripe texture. Fully coupled simulations further demonstrate phonon-mediated excitation of a remote YIG layer and show that efficient propagating-phonon generation requires spatially asymmetric magnon modes, establishing magnetic texture as a control parameter for magnon--phonon coupling.

\end{abstract}

\maketitle

\section{Introduction}

Magnetoelastic coupling in low-damping magnetic insulators provides a route to control spin dynamics through long-lived lattice excitations. Magnons are considered as promising carriers for spin angular momentum~\cite{Flebus_IOP2024}, while phonons offer long lifetimes~\cite{dutoit1974microwave}, efficient confinement in acoustic cavities~\cite{Schlitz2020}, and long-range transport of spin angular momentum~\cite{an2020coherent,ruckriegel2020long,cornelissen2015long}. Their mutual coupling therefore underpins a broad class of hybrid magnonic phenomena, including coherent transduction, phonon-assisted signal processing, and phonon-mediated angular-momentum transfer~\cite{lachance2019hybrid,an2020coherent,bienfait2019phonon,hioki2022coherent}. Yttrium iron garnet (YIG) is a suitable material platform for such studies because of its low magnetic damping properties~\cite{chumak2015magnon,spencer1959low,serga2010yig,cherepanov1993saga} that allow one to readily access nonlinear spin-wave interactions for novel spintronic and magnonic functionalities~\cite{Kajiwara_Nature2010, Kurebayashi_NatMater2011, Lee_PRL2023, Makiuchi_NatMater2024}. Gadolinium gallium garnet (GGG) is the substrate of choice for YIG, supporting high-quality epitaxial growth~\cite{Dubs2020lpe}. Its low ultrasonic attenuation~\cite{dutoit1974microwave}, long phonon mean free path~\cite{spencer1962temperature}, and favorable acoustic impedance matching with YIG~\cite{polzikova2019acoustic} make YIG/GGG heterostructures well suited for magnon--phonon coupling studies. In such heterostructures, spin-waves (magnons) can couple to lattice vibrations (phonons) through the strain dependence of the magnetocrystalline anisotropy energy~\cite{kittel1958interaction,bommel1959excitation,dreher2012surface,zhang2016cavity}.

Recent studies have established YIG/GGG heterostructures as a versatile platform for magnetoelastic phenomena, ranging from hybrid magnon--phonon quasiparticles (magnon polarons) formed near spin-wave and acoustic-wave avoided crossings~\cite{an2020coherent, Schlitz2020,hioki2022coherent,kikkawa2016magnon,cornelissen2017nonlocal} to application-oriented functionalities such as microwave acoustic transducers~\cite{pomerantz1961excitation,reeder2003characteristics},
parametric acoustic oscillators~\cite{chowdhury2015parametric}, and nonreciprocal acoustic-wave control~\cite{popa2014non,matthews1962acoustic}. In YIG, this coupling has also been shown to influence spin transport~\cite{comstock1963generation,kikkawa2016magnon,cornelissen2017nonlocal}, angular-momentum transfer~\cite{garanin2015angular}, and phonon pumping into adjacent dielectrics, which enhances magnetic damping~\cite{streib2018damping,schlitz2022magnetization}. For YIG/GGG stacks specifically, standing acoustic modes extending across the heterostructure can hybridize with magnetic resonances~\cite{Schlitz2020,an2020coherent,man2017direct}, generate phonon-pumping signatures in ferromagnetic resonance~\cite{sato2021dynamic,hioki2022coherent}, and form dense multimode phonon spectra with MHz-scale free spectral range, where the achievable coupling is governed by mode overlap, thickness matching, and cavity-phonon confinement~\cite{an2023optimizing,xu2021coherent}. However, these studies have focused primarily on nearly uniform magnon modes or macrospin-like resonances with clearly resolved hybridization features. Stripe-domain states in garnet films have been studied in the context of domain-wall resonances and bubble domains~\cite{HubertSchafer1998}, but the magnetoelastic
response of the nonuniform stripe-domain magnon spectrum has not been systematically investigated.

In this work, we investigate magnetoelastic coupling in a \SI{3}{\micro\meter}-thick YIG film on a \SI{0.5}{\milli\meter}-thick GGG substrate in the low-field stripe-domain regime. Recent work~\cite{Prestwood2025StripeYIG} showed that YIG films above a critical thickness support a stripe-domain ground state with rich spin-wave modes, providing the basis for the present study. Using broadband ferromagnetic resonance (FMR), spectral fitting, two-dimensional overlap analysis, and fully coupled finite-element simulations, we show that spin-wave modes excited in these stripe domains are weakly coupled to confined thickness-shear acoustic modes of the GGG substrate. This coupling appears experimentally as a field-independent phonon-comb modulation with a free spectral range of approximately \SI{3.5}{\mega\hertz}, rather than as resolved avoided crossings. The extracted coupling rate is lower than the relaxation rate of magnons, placing the system in the weak-coupling regime. Finite-element mode overlap calculations show that the weak coupling is not caused by a negligible local magnetoelastic interaction, but by strong coherent cancellation from the phase structure and sign reversal of the stripe-domain magnetization. Fully coupled YIG/GGG/YIG simulations further demonstrate that stripe-domain excitation can launch shear phonon waves that mediate dynamical coupling of two magnetic layers across the substrate. These results establish magnetic texture as a control parameter for phonon-mediated magnon coupling in insulating magnetic heterostructures.

\section{Methods}

\begin{figure}
    \centering
    \includegraphics[width=1\linewidth]{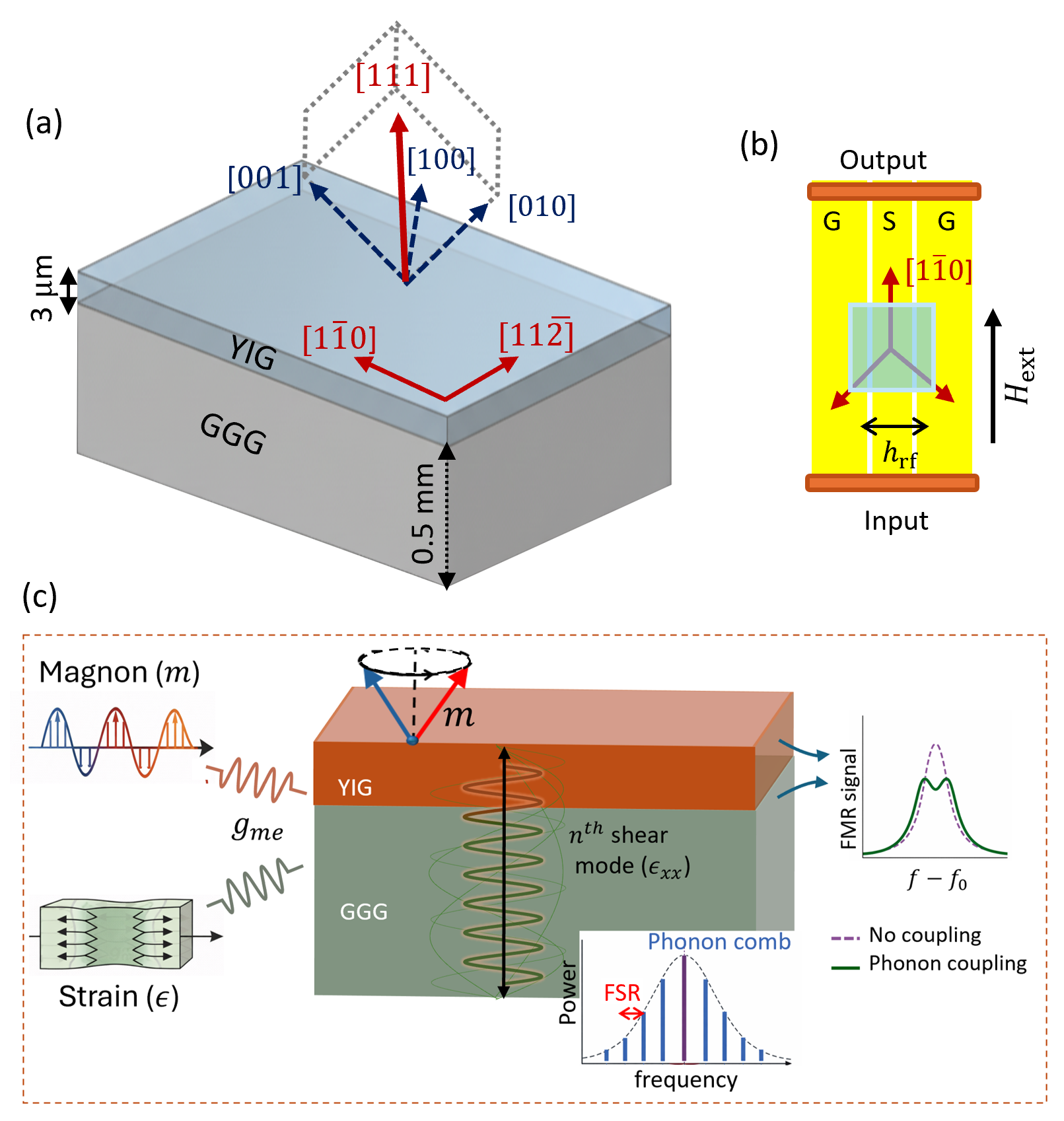}
    \caption{(a) Schematic of a \SI{3}{\micro\meter}-thick YIG thin film epitaxially grown on a GGG substrate oriented along the $[111]$ direction. (b) Schematic of the FMR measurement geometry, showing the static external magnetic field applied in the film plane along the $[1\bar{1}0]$ direction. (c) Conceptual illustration of magnetoelastic coupling in the YIG/GGG heterostructure: microwave-driven magnons in YIG generate strain and excite confined thickness-shear phonon modes in GGG, forming a dense phonon comb. The phonons then
    couple back to the magnons and perturb the FMR response.}
    \label{fig:Intro}
\end{figure}

The investigated sample comprises a \SI{3}{\micro\meter}-thick YIG film grown by liquid-phase epitaxy on a \SI{500}{\micro\meter}-thick GGG substrate, with the film normal along the crystallographic $[111]$ direction [Fig.~\ref{fig:Intro}(a)]. At this thickness, the film supports a stripe-domain state at a low in-plane magnetic field, stabilized by the competition between weak perpendicular magnetic anisotropy and the film shape anisotropy~\cite{HubertSchafer1998, Prestwood2025StripeYIG}. FMR spectra were measured at room temperature using a coplanar waveguide (CPW) and a vector network analyzer in a flip-chip geometry.  The complex transmission coefficient $S_{21}$ was recorded over $0$--$2\,\mathrm{GHz}$ with a frequency step of $0.04\,\mathrm{MHz}$ and over $0$--$10\,\mathrm{mT}$ with a field step of $0.044\,\mathrm{mT}$. The external magnetic field ($H_{\mathrm{ext}}$) was applied along $[1\bar{1}0]$, while the microwave excitation field was applied along $[11\bar{2}]$, as shown in Fig.~\ref{fig:Intro}(b).

The magnetoelastic response is analyzed in the weak-coupling limit, where a stripe-domain magnon mode interacts with nearby confined thickness-shear acoustic modes of the YIG/GGG heterostructure. Using a coupled harmonic-oscillator model~\cite{Herskind_NatPhys2009,an2020coherent,Khan_PRB2021,Zollitsch_NComm2023}, the local linear response is written as
\begin{equation}
\begin{split}
\left(f-f_{\mathrm{m},i}+i\kappa_{\mathrm{m},i}\right)m_i
&=
h_{\mathrm{rf}}+\sum_n g_{i,n}u_n, \\
\left(f-f_{\mathrm{ph},n}
+i\kappa_{\mathrm{ph},n}\right)u_n
&=
g_{i,n}m_i .
\end{split}
\label{eq:coupled_oscillator}
\end{equation}
Here, \(f\) is the microwave-drive frequency, while \(f_{\mathrm{m},i}\) and \(f_{\mathrm{ph},n}\) are the resonance frequencies of magnon mode \(i\) and phonon mode \(n\), respectively. The complex amplitudes \(m_i\) and \(u_n\) describe the corresponding dynamic magnetization and elastic displacement responses. The quantities \(\kappa_{\mathrm{m},i}\) and \(\kappa_{\mathrm{ph},n}\) are the half-widths at half-maximum (HWHM) of the magnon and phonon resonances, respectively, expressed in frequency units. The mode-dependent magnetoelastic coupling rate between magnon mode \(i\) and phonon mode \(n\) is denoted by \(g_{i,n}\), ensuring reciprocal coupling between the magnetic and elastic subsystems. The term \(h_{\mathrm{rf}}\) is the complex amplitude of the microwave magnetic field that directly drives the magnon mode, and the summation over \(n\) accounts for its coupling to the set of nearby confined thickness-shear phonon modes. For fitting individual peak--dip features, the response is treated locally as one broad magnon branch coupled to one nearby phonon resonance, while unresolved magnon modes and inhomogeneous broadening enter through the effective magnon linewidth and background lineshape. The physical picture underlying this model (microwave-driven magnons generating shear strain that excites confined phonon modes, which in turn perturb the FMR response) is illustrated in Fig.~\ref{fig:Intro}(c).

Finite-element simulations were performed in \textsc{COMSOL Multiphysics} v.~6.2 using the frequency-domain Micromagnetics Module coupled to the Solid Mechanics module~\cite{COMSOL62,zhang2023frequency}. The calculation solves for complex dynamic magnetization $\delta\mathbf{m}$ about the relaxed stripe-domain equilibrium state $\mathbf{m}_0$, together with the complex elastic displacement field $\mathbf{u}$. Magnetoelastic coupling was introduced through the frequency-domain magnetoelastic field~\cite{kittel1958interaction}
\begin{equation}
H_{\mathrm{me},i}
=
-\frac{2B_{\mathrm{1}}}{\mu_0M_{\mathrm{s}}}
m_{0i}\epsilon_{ii}
-
\frac{2B_{\mathrm{2}}}{\mu_0M_{\mathrm{s}}}
\sum_{j\ne i}
m_{0j}\epsilon_{ij},
\label{eq:Hme_compact}
\end{equation}
where $i,j\in\{x,y,z\}$, $B_{\mathrm{1}}$ and $B_{\mathrm{2}}$ are the cubic magnetoelastic coefficients, $M_{\mathrm{s}}$ is the saturation magnetization, and $\epsilon_{ij}$ are the tensor strain components of the dynamic displacement field. The simulations used the following YIG parameters~\cite{Prestwood2025StripeYIG, an2020coherent}: saturation magnetization $M_{\mathrm{s}}=\SI{1.96e5}{\ampere\per\metre}$, first cubic anisotropy constant $K_1=\SI{-610}{\joule\per\metre\cubed}$, uniaxial anisotropy parameter $K_{\mathrm{u}}=\SI{1000}{\joule\per\metre\cubed}$, exchange coefficient $A_{\mathrm{ex}}=\SI{6.5e-12}{\joule\per\metre}$, and magnetoelastic constants  $B_{\mathrm{1}} = \SI{3.5e5}{\joule\per\metre\cubed}$, $B_{\mathrm{2}} = \SI{7.0e5}{\joule\per\metre\cubed}$.
\begin{figure*}
    \centering
    \includegraphics[width=1\linewidth]{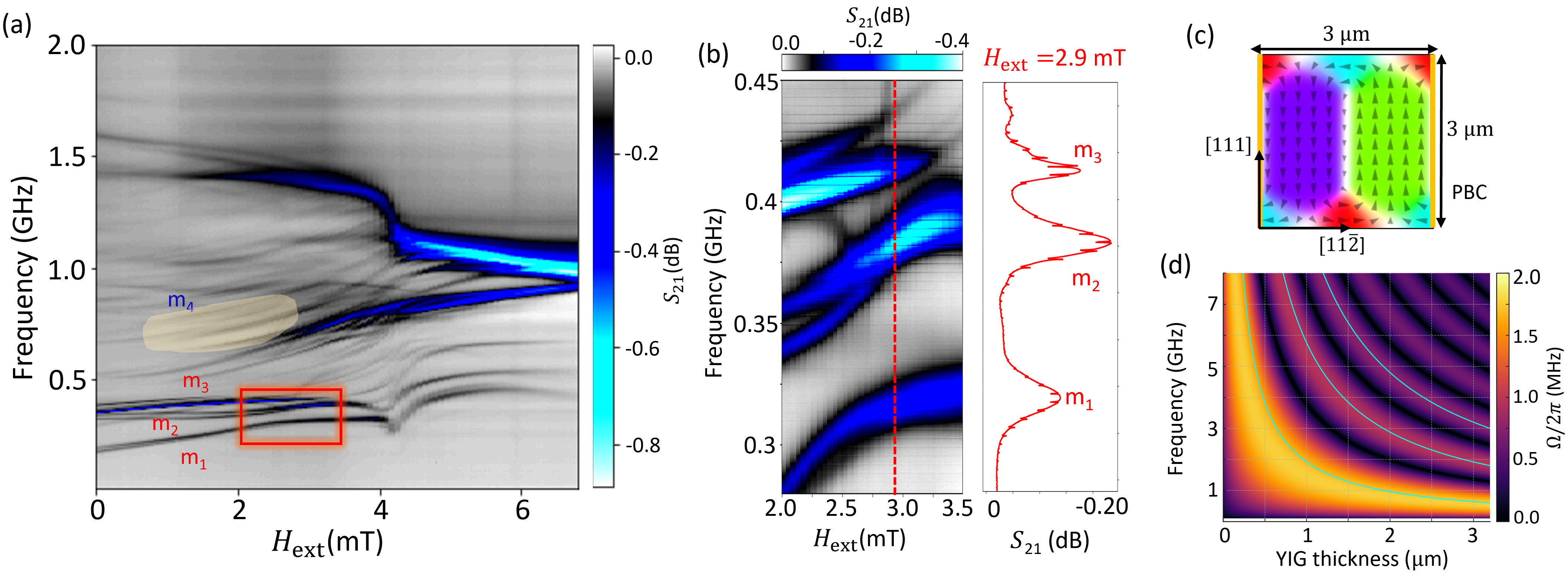}
    \caption{(a) Field--frequency microwave transmission spectrum ($S_{21}$) of a
    YIG(\SI{3}{\micro\meter})/GGG heterostructure measured using broadband CPW-FMR. Multiple magnon branches are labeled $\mathrm{m_1}$--$\mathrm{m_4}$. (b) Zoomed microwave transmission spectral map for frequencies below \SI{0.5}{\giga\hertz} and a magnetic-field range \SI{2.0}{}--\SI{3.5}{\milli\tesla}. The transmission spectrum at a representative field of \SI{2.9}{\milli\tesla} (red vertical line in two-dimensional spectral map) is shown on the right to highlight the periodic phonon modulation. (c) Micromagnetically simulated remanent magnetization state of the YIG film, computed in a \SI{3}{\micro\meter} $\times$ \SI{3}{\micro\meter} cross-sectional cell with periodic boundary conditions along the lateral ([$11\bar{2}$]) direction, revealing a stripe periodicity of $\approx$\,\SI{3}{\micro\meter}. (d) Calculated magnetoelastic coupling strength as a function of YIG film thickness and magnon frequency, showing a maximum near \SI{0.5}{\giga\hertz} for a YIG thickness of \SI{3}{\micro\meter}.}
    \label{fig:map}
\end{figure*}

The reciprocal magnon-to-phonon coupling was introduced in Solid Mechanics through a magnetoelastic body load~\cite{yamamoto2020non} in the YIG domains and a boundary load~\cite{sato2021dynamic} at each YIG/GGG interface. The body-load component along a fixed direction $i$ is
\begin{equation}
F^{\mathrm{me}}_{i}
=
2B_{\mathrm{1}}\,\partial_i
\left(
m_{0i}\delta m_i
\right)
+
B_{\mathrm{2}}
\sum_{j\ne i}
\partial_j
\left(
m_{0j}\delta m_i
+
m_{0i}\delta m_j
\right),
\label{eq:bodyload_compact}
\end{equation}
where $\delta m_i$ is the complex dynamic magnetization component and $i$ is a fixed, unsummed component index.
The corresponding boundary load is
\begin{equation}
T^{\mathrm{me}}_{i}
=
-
\left[
2B_{\mathrm{1}}m_{0i}\delta m_i n_i
+
B_{\mathrm{2}}
\sum_{j\ne i}
\left(
m_{0i}\delta m_j
+
\delta m_i m_{0j}
\right)n_j
\right].
\label{eq:boundaryload_compact}
\end{equation}
Here $\mathbf{n}_{\mathrm{int}}=(n_x,n_y,n_z)$ is the interface normal, defined at each YIG/GGG interface to point away from the YIG layer and into the GGG spacer. 

The simulations were performed in a two-dimensional $x$--$y$ cross-section with $\hat{x}\parallel[11\bar{2}]$ along the stripe modulation direction and $\hat{y}\parallel[111]$ along the film thickness; the stripe domains extend along
$\hat{z}\parallel[1\bar{1}0]$.
Accordingly, only the displacement components and derivatives resolved in the simulated cross-section are retained in Eqs.~\eqref{eq:Hme_compact}--\eqref{eq:boundaryload_compact}. Periodic boundary conditions (PBC) were applied laterally, and the mesh was refined in the YIG and near the YIG/GGG interface to resolve the domain texture and the shortest relevant shear-acoustic wavelength. All frequency-domain calculations used a small harmonic drive to remain in the linear-response regime.

\section{Results and Discussion}

Figure~\ref{fig:map}(a) shows the field--frequency microwave transmission spectrum of the YIG/GGG film measured with $H_{\mathrm{ext}}\parallel[1\bar{1}0]$. The measurements were performed below saturation, where the YIG film exhibits a stripe-domain state~\cite{Prestwood2025StripeYIG}. Its spatially nonuniform states host both domain-wall edge and bulk modes, including their hybridized and high-order dynamics~\cite{Prestwood2025StripeYIG}, as shown in Fig.~\ref{fig:map}(a).

The spectrum can be separated into two field regimes. At lower fields, up to approximately $H_{\mathrm{ext}}\approx\SI{4.1}{\milli\tesla}$, several well-resolved low-frequency branches are present, corresponding to the stable stripe-domain regime.
Above this field, the stripe texture becomes distorted as the magnetization tilts toward the in-plane crystallographic easy directions~\cite{Prestwood2025StripeYIG}. The spectrum follows this rotation, and two dominant branches in the multi-mode response merge near \SI{1}{\giga\hertz} with increasing field. The transition toward a uniform magnetic state occurs near
$H_{\mathrm{sat}}\approx\SI{7}{\milli\tesla}$, beyond which a single Kittel-like resonance is observed. This field-driven transformation is qualitatively reproduced by micromagnetic simulations using the experimentally determined magnetic parameters
(Supplementary Fig.~S1). Figure~\ref{fig:map}(c) shows the simulated thickness profile of the remanent magnetization in YIG, revealing a lateral stripe periodicity of approximately \SI{3}{\micro\meter}, consistent with the nonuniform magnetic state observed in the measured spectrum.

Within the low-field stripe-domain regime, additional fine structure appears on the low-frequency modes labeled $\mathrm{m_1}$, $\mathrm{m_2}$, and $\mathrm{m_3}$ in Fig.~\ref{fig:map}(a). Figure~\ref{fig:map}(b) shows a magnified view over \SIrange{2.0}{3.5}{\milli\tesla} and frequencies below \SI{0.5}{\giga\hertz}. The fine structure appears as weak, periodic, field-independent lines crossing the magnon modes. A representative line cut at $H_{\mathrm{ext}}=\SI{2.9}{\milli\tesla}$ highlights the corresponding periodic modulations superimposed on the broader magnon resonances. Their field-independent character indicates an acoustic rather than magnetic origin. The measured spacing is $\Delta f\approx\SI{3.54}{\mega\hertz}$, matching the free spectral range (FSR) of standing thickness-shear modes in the YIG/GGG heterostructure. For YIG thickness $d$ and GGG thickness $s$, the acoustic mode spacing is $\Delta f_{\mathrm{ph}}=v_{\mathrm{T}}/[2(d+s)]$, where $v_{\mathrm{T}}$ is the transverse sound velocity along $[111]$. Using $v_{\mathrm{T}}=\SI{3.53e3}{\metre\per\second}$ for GGG along $[111]$~\cite{spencer1963microwave} gives $\Delta f_{\mathrm{ph}}\approx\SI{3.51}{\mega\hertz}$, in close agreement with the experimentally observed modulation.

The phonon comb is most visible on the low-frequency stripe-domain modes below $\sim\SI{0.5}{\giga\hertz}$, indicating that its visibility is governed not only by the acoustic mode density but also by the magnetoelastic overlap between the magnon mode and the thickness-shear strain field. To estimate the relevant frequency range, we calculate the thickness-dependent coupling strength for a shear acoustic cavity coupled to a magnetic film. Following An~et~al.~\cite{an2020coherent,an2023optimizing}, the coupling scale for an idealized uniformly out-of-plane magnetized film is
\begin{equation}
\Omega_{[111]} =
2\bar{B}_{[111]}
\sqrt{\frac{\gamma}{\omega \rho M_{\mathrm{s}} d(d+s)}}
\left(1-\cos\frac{n\pi d}{d+s}\right),
\label{eq:coupling_scale}
\end{equation}
where $\bar{B}_{[111]}=(B_{\mathrm{2}}+2B_{\mathrm{1}})/3$, $\rho$ is the YIG mass density, $M_{\mathrm{s}}$ is the saturation magnetization, $d$ and $s$ are the YIG and GGG thicknesses, and $\gamma$ is the gyromagnetic ratio. The calculated coupling strength for the present structure is shown in Fig.~\ref{fig:map}(d).

The calculation shows enhanced coupling when the shear-acoustic wavelength satisfies an approximate half-wavelength condition across the YIG layer, $\lambda_n\approx 2d$. For the present $d=\SI{3}{\micro\meter}$ film, this condition places the largest
coupling near $\sim\SI{0.5}{\giga\hertz}$, matching the frequency range where the periodic modulation is observed. This provides a thickness-matching explanation for why the phonon comb is most visible on the low-frequency magnon modes. However, 
Eq.~\eqref{eq:coupling_scale} assumes a uniformly magnetized out-of-plane film, whereas the experiment is performed in a stripe-domain state. The value in Fig.~\ref{fig:map}(d) should therefore be interpreted as an upper limit on the magnetoelastic coupling strength; the actual coupling depends on the spatial overlap between the stripe-domain magnon modes and the shear strain, including constructive and destructive contributions from different domains and domain-wall regions.

\begin{figure}[t]
  \centering
  \includegraphics[width=\linewidth]{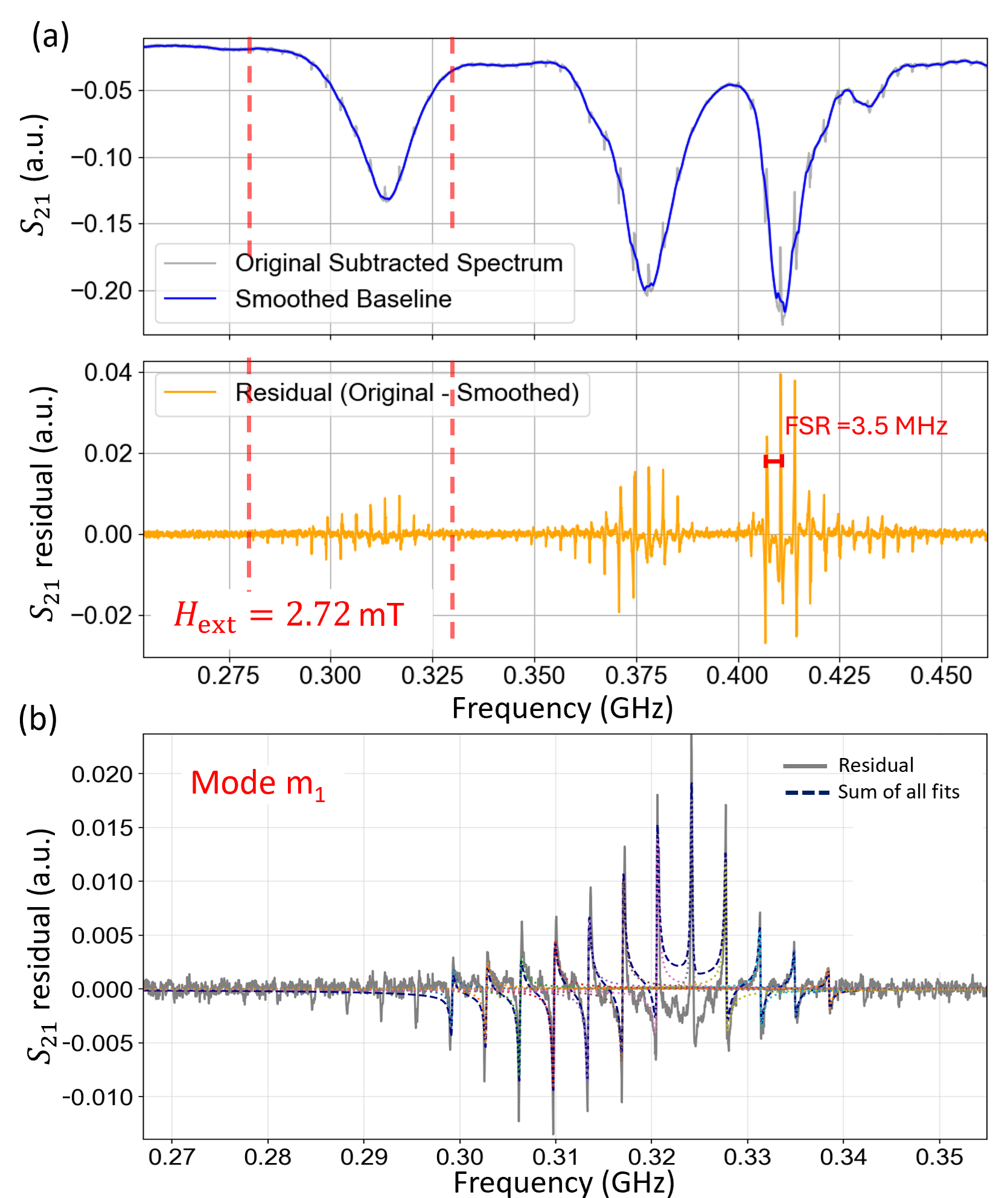}
  \caption{(a) Representative frequency sweep at fixed magnetic field, showing the raw spectrum together with a smooth background
  representing the broad magnon response of modes $\mathrm{m_1}$-$\mathrm{m_4}$. (b) Residual spectrum obtained after background subtraction, revealing narrow, periodically spaced peak--dip features with free spectral range $\Delta f\approx\SI{3.5}{\mega\hertz}$.}
  \label{fig:residual}
\end{figure}

To quantify this weak, overlap-limited interaction, we next analyze the phonon-comb features superimposed on individual stripe-domain magnon modes. Figure~\ref{fig:residual}(a) shows a representative frequency sweep at $H_{\mathrm{ext}}=\SI{2.72}{\milli\tesla}$. The slowly varying magnon spectrum was obtained using a Savitzky--Golay smoothing procedure and subtracted from the measured spectrum in order to discuss the phononic modes. The residual spectrum in Fig.~\ref{fig:residual}(b) reveals narrow peak--dip features with periodic spacing $\Delta f\approx\SI{3.54}{\mega\hertz}$, consistent with the thickness-shear acoustic FSR identified in Fig.~\ref{fig:map}.

The effective magnon linewidths were extracted from the smoothed magnon resonances using either a Fano-type~\cite{Miroshnichenko2010fano, Li2022fano} or mixed Lorentzian line shape~\cite{Harder2011lineshape}, depending on the peak symmetry. The mixed Lorentzian form was used for nearly isolated resonances where the asymmetry mainly reflects absorptive--dispersive phase mixing, whereas the Fano form was used for strongly asymmetric resonances affected by a slowly varying background or nearby unresolved magnetic response. At $H_{\mathrm{ext}}=\SI{2.72}{\milli\tesla}$, the extracted effective linewidths $\kappa_{\mathrm{m},i}$ are $\SI{8.99}{\mega\hertz}$, $\SI{8.39}{\mega\hertz}$, and $\SI{5.84}{\mega\hertz}$ for modes $\mathrm{m_1}$, $\mathrm{m_2}$, and $\mathrm{m_3}$, respectively. Note that these are effective linewidths and do not represent intrinsic relaxation rates, because they include internal-field inhomogeneity, domain and domain-wall nonuniformity, mode mixing, and unresolved overlap of nearby spin-wave modes.

The phonon linewidth was determined independently by fitting isolated phonon resonances away from the magnon resonance with a Lorentzian, yielding $\kappa_{\mathrm{ph}}\approx\SI{0.14}{\mega\hertz}$. This corresponds to a phonon lifetime
$\tau_{\mathrm{ph}}\approx(2\pi\kappa_{\mathrm{ph}})^{-1}\approx \SI{1.1}{\micro\second}$. For phonon frequencies from \SIrange{0.32}{0.5}{\giga\hertz}, the corresponding quality factor is $Q=f_{\mathrm{ph}}/\kappa_{\mathrm{ph}}\approx\numrange{2.2e3}{3.5e3}$. With $v_\mathrm{T}\approx\SI{3.53e3}{\metre\per\second}$, the acoustic wave travels $v_\mathrm{T}\tau_{\mathrm{ph}}\approx\SI{3.9}{\milli\meter}$ before decaying, corresponding to roughly four round trips across the $d+s\approx\SI{0.503}{\milli\meter}$ YIG/GGG acoustic cavity.

The peak--dip structures within the magnon envelope were fitted using a mixed Lorentzian form,
\begin{equation}
S_{21}(f) = a\frac{\Gamma^2}{(f-f_0)^2+\Gamma^2}+b\frac{\Gamma (f-f_0)}{(f-f_0)^2+\Gamma^2},
\label{eq:mixed_lorentzian}
\end{equation}
where $f_0$ is the center frequency, $\Gamma$ is the half-width parameter, and $a$ and $b$ describe the absorptive and dispersive components. The coupling rate $g_{i,n}$ to each phonon mode was determined from the amplitude of the fitted peak-dip feature relative to the magnon background. Since $g_{i,n} < \kappa_{\mathrm{m}}$ for all analyzed modes, the system lies in the weak-coupling regime and no avoided crossing is resolved; the extracted values are summarized in Table~\ref{tab:coupling}. The interaction is therefore limited by the broad stripe-domain magnon resonances rather than by phonon loss: the phonons are long-lived, but magnon coherence is lost before resolved coherent exchange can develop.

Using the extracted $g_{i,n}$, $\kappa_{\mathrm{m}}$, and $\kappa_{\mathrm{ph}}$, the cooperativity is calculated as $C_{i,n}=g_{i,n}^{2}/(\kappa_{\mathrm{m}}\kappa_{\mathrm{ph}})$, following Refs.~\cite{an2020coherent,an2023optimizing}.
The extracted parameters are summarized in Table~\ref{tab:coupling}. The mean cooperativity remains below unity for all analyzed modes, confirming weak magnetoelastic coupling between the stripe-domain magnons and confined shear-acoustic modes. Mode $\mathrm{m_3}$ shows the largest mean cooperativity, followed by $\mathrm{m_2}$ and $\mathrm{m_1}$, but all values remain well below unity, explaining why the experiment shows periodic modulation of the magnon resonance rather than resolved avoided crossings.

\begin{table}
\centering
\caption{Extracted half-linewidths, coupling rate $g$, and mean cooperativity $C$
(with fit uncertainties) for representative stripe-domain magnon modes coupled to the
shear-acoustic phonon comb at $H_{\mathrm{ext}}=\SI{2.72}{\milli\tesla}$.}
\begin{tabular}{ccccc}
\hline\hline
Mode & $\kappa_{\mathrm{ph}}$ & $\kappa_{\mathrm{m}}$ & $g$ & $C$ \\
     & (MHz) & (MHz) & (MHz) & \\
\hline
$m_1$ & 0.14 & 8.99 & $0.33\pm0.01$ & $0.083\pm0.005$ \\
$m_2$ & 0.14 & 8.39 & $0.40\pm0.04$ & $0.140\pm0.032$ \\
$m_3$ & 0.14 & 5.84 & $0.54\pm0.03$ & $0.348\pm0.040$ \\
\hline\hline
\end{tabular}
\label{tab:coupling}
\end{table}

\begin{figure*}[t]
    \centering
    \includegraphics[width=\linewidth]{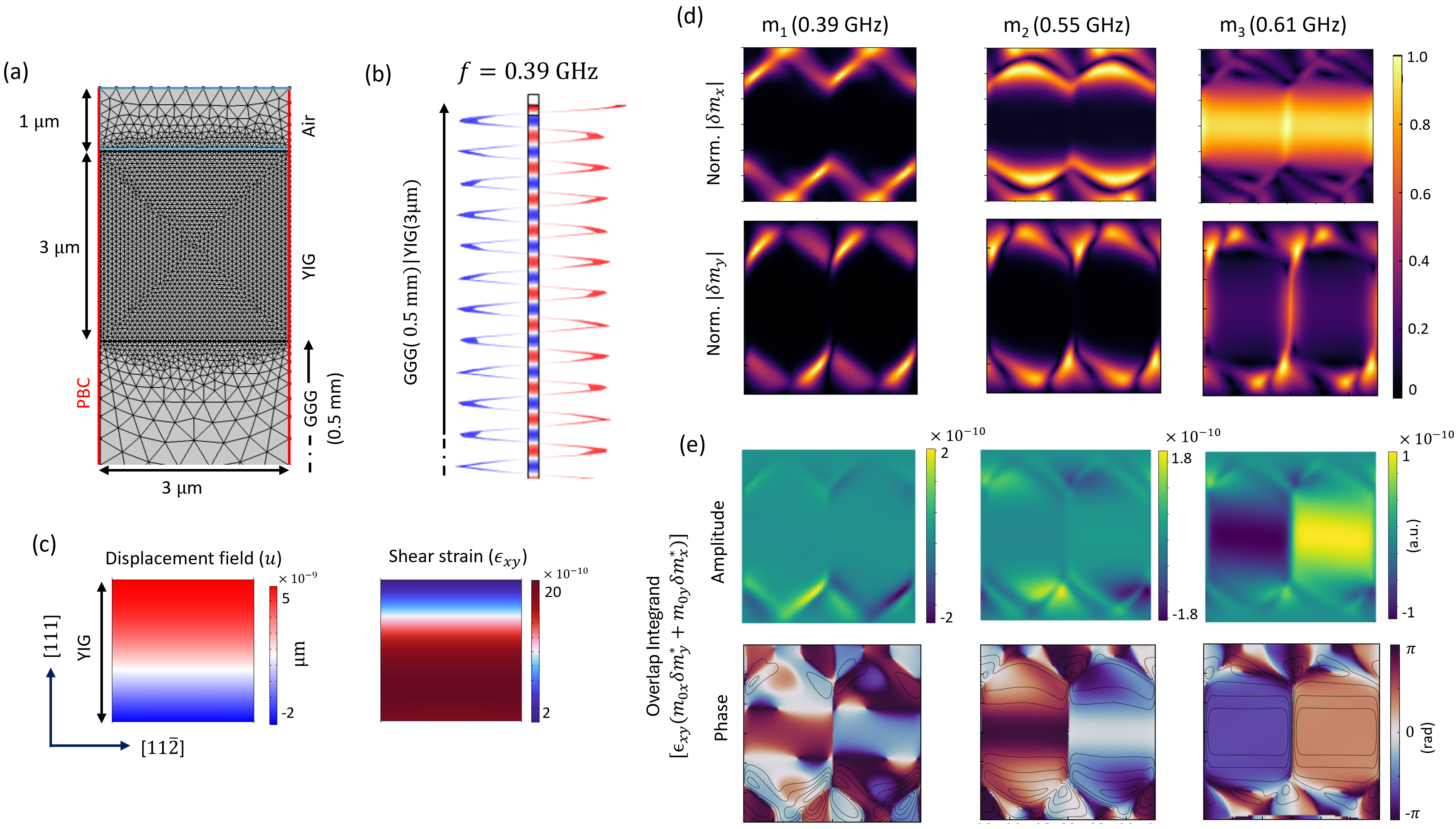}
    \caption{
    Finite-element and micromagnetic analysis of the projected magnetoelastic overlap in the stripe-domain YIG/GGG heterostructure. (a) Meshed simulation geometry consisting of a \SI{3}{\micro\meter} YIG film on a \SI{0.5}{\milli\meter} GGG substrate. Periodic boundary conditions are applied along the lateral edges, and an air domain is included above the YIG layer to account for the magnetic stray field. (b) Spatial displacement field of the YIG/GGG heterostructure at a representative phonon frequency, driven by a magnetoelastic body load in the YIG and a boundary load at the YIG/GGG interface. The scaled deformation illustrates the distribution of $u_x$ and $u_y$ across the heterostructure, confirming a thickness-shear mode character. (c) The zoomed panels show the $x$-component of displacement, $u_x$, and the corresponding shear strain, $\epsilon_{xy}$, in the YIG layer. (d) Simulated dynamic magnetization profiles for the low-frequency stripe-domain modes $\mathrm{m_1}$--$\mathrm{m_3}$, shown using the normalized dynamic components $|\delta m_x|$ (top panel) and $|\delta m_y|$ (bottom panel). (e) Amplitude and phase of the simulated two-dimensional magnetoelastic overlap
    integrand, $\mathcal{I}_{xy,i}^{2\mathrm{D}}$, for modes $\mathrm{m_1}$--$\mathrm{m_3}$.
    }
    \label{fig:simulation_results}
\end{figure*}

\begin{figure*}
    \centering
    \includegraphics[width=0.9\linewidth]{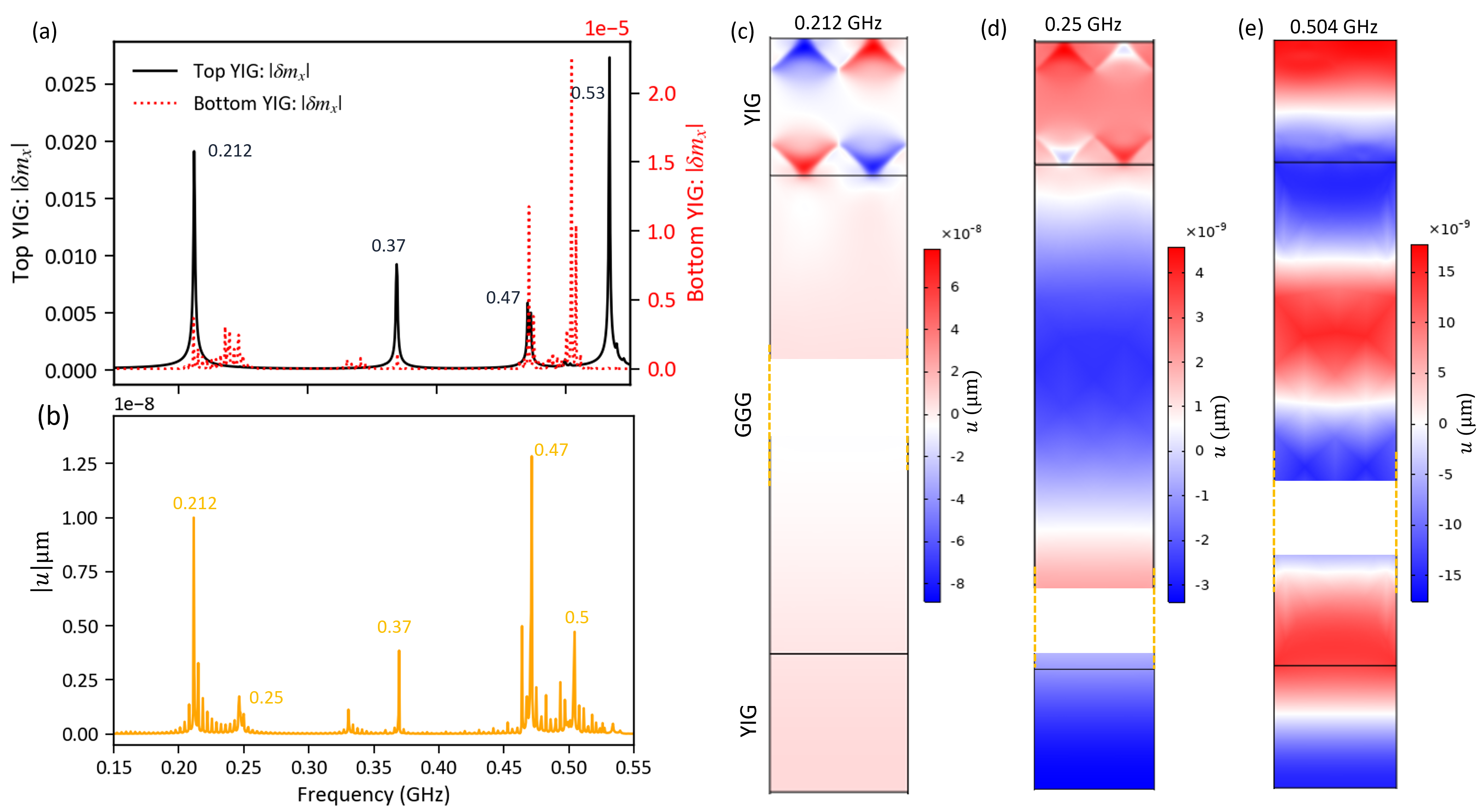}
    \caption{
   Fully coupled magnetoelastic simulation of phonon-mediated magnon transfer in a YIG/GGG/YIG structure.
    (a) Simulated dynamic magnetization spectra, obtained from the Fourier amplitude of $\delta m_x$, for the rf-driven top YIG layer and the indirectly excited bottom YIG layer via phonons. (b) Corresponding phonon spectrum generated by the driven top-layer magnon dynamics, showing several shear-cavity modes within the magnetic resonance bandwidth. (c)--(e) Displacement-field profiles $u_x$ of the coupled YIG/GGG/YIG system at representative frequencies (0.21, 0.25, 0.5\,GHz).}
    \label{fig:fig5}
\end{figure*}

To identify the spatial origin of this weak and mode-dependent coupling, we evaluated the magnetoelastic overlap between the simulated stripe-domain magnon modes and the shear-acoustic strain field. In a three-dimensional description, the coupling rate is set by the linearized magnetoelastic interaction energy,
\begin{equation}
g_{i,n} = \frac{|U_{i,n}^{\mathrm{me}}|}{h}, 
\quad
U_{i,n}^{\mathrm{me}}
=
\int_{V_{\mathrm{YIG}}}
\delta f_{\mathrm{me},i,n}(\mathbf{r})\,dV,
\label{eq:overlap_coupling}
\end{equation} 
where $g_{i,n}$ is the coupling rate in frequency units, $h$ is Planck's constant, $U_{i,n}^{\mathrm{me}}$ is the magnetoelastic interaction energy between magnon mode $i$ and phonon mode $n$, and $\delta f_{\mathrm{me},i,n}$ is the corresponding linearized magnetoelastic energy density.

For the shear part of the cubic magnetoelastic energy density, 
\begin{equation}
f_{\mathrm{me}}^{\mathrm{shear}}
=
2B_{\mathrm{2}}
\left(
\epsilon_{xy}m_xm_y+
\epsilon_{yz}m_ym_z+
\epsilon_{zx}m_zm_x
\right),
\end{equation}
where $B_{\mathrm{2}}$ is the shear magnetoelastic coefficient, $\epsilon_{\alpha\beta}$ are the tensor shear-strain components, and $m_{\alpha}$ are components of the normalized magnetization. Linearizing about the stripe-domain equilibrium state,
$\mathbf{m}=\mathbf{m}_0+\delta\mathbf{m}_i$, gives
\begin{equation}
\begin{split}
\delta f_{\mathrm{me},i}^{\mathrm{shear}}
=
2B_{\mathrm{2}}
\Big[
&\epsilon_{xy,n}
\left(
m_{0x}\delta m_{y,i}^{*}
+
m_{0y}\delta m_{x,i}^{*}
\right) \\
+
&\epsilon_{yz,n}
\left(
m_{0y}\delta m_{z,i}^{*}
+
m_{0z}\delta m_{y,i}^{*}
\right) \\
+
&\epsilon_{zx,n}
\left(
m_{0z}\delta m_{x,i}^{*}
+
m_{0x}\delta m_{z,i}^{*}
\right)
\Big].
\end{split}
\end{equation}
Since $\mathbf{m}$ is a unit vector, no additional factor of $M_{\mathrm{s}}$ appears in the overlap expression.

The simulations were performed in the $x$--$y$ cross-section, with $\hat{x}\parallel[11\bar{2}]$ along the stripe modulation direction and $\hat{y}\parallel[111]$ along the film thickness; the stripe domains extend along $\hat{z}\parallel[1\bar{1}0]$. The two-dimensional model captures the lateral stripe modulation and the thickness-dependent shear strain, but excludes displacement along the stripe direction. The resolved contribution is therefore the projected $\epsilon_{xy}$ channel,
\begin{equation}
\begin{split}
\mathcal{I}_{xy,i}^{2\mathrm{D}}(x,y)
&=
\epsilon_{xy,n}(x,y)
\Big[
m_{0x}(x,y)\delta m_{y,i}^{*}(x,y) \\
&\quad+
m_{0y}(x,y)\delta m_{x,i}^{*}(x,y)
\Big],
\end{split}
\end{equation}
where $\mathcal{I}_{xy,i}^{2\mathrm{D}}$ is the projected local overlap integrand in the simulated cross-section, apart from the constant prefactor $2B_{\mathrm{2}}$. The corresponding coherent projected overlap is
\begin{equation}
I_{xy,i}^{2\mathrm{D}}
=
\int_{A_{\mathrm{YIG}}}
\mathcal{I}_{xy,i}^{2\mathrm{D}}(x,y)\,dA .
\end{equation}
The absolute value is taken only after integration, so that phase cancellation between domains and domain-wall regions is retained. Because the calculation is two-dimensional and omits the remaining shear channels, $I_{xy,i}^{2\mathrm{D}}$ is used as a relative projected overlap metric; the absolute coupling strength is obtained from the experimentally measured peak--dip splitting.

The simulation geometry is shown in Fig.~\ref{fig:simulation_results}(a), with the meshed YIG/GGG cross-section, PBC along the lateral edges, and an air domain above the YIG layer. To verify that the dominant elastic response is a thickness-shear mode, Fig.~\ref{fig:simulation_results}(b) shows the displacement field across the YIG/GGG heterostructure at a representative frequency, with the scaled deformation illustrating the spatial distribution of $u_x$ and $u_y$. The zoomed panels in Fig.~\ref{fig:simulation_results}(c) confirm that the $x$-component of displacement $u_x$ and the corresponding shear strain $\epsilon_{xy}$ are concentrated within the YIG layer, establishing the strain field that enters the overlap integral.

Figure~\ref{fig:simulation_results}(d) shows the simulated dynamic magnetization profiles of the three low-frequency modes analyzed experimentally. Mode $\mathrm{m_1}$ is localized mainly near the domain walls and flux-closure regions, where the magnetization rotates between neighboring stripe domains. Mode $\mathrm{m_2}$ has stronger weight near the upper and lower tilted regions of the stripe domains, close to the flux-closure caps. Mode $\mathrm{m_3}$ is more extended across the domain interior and corresponds to a bulk-like stripe mode. The amplitude and phase maps of $\mathcal{I}_{xy,i}^{2\mathrm{D}}$ in Fig.~\ref{fig:simulation_results}(e) show which regions add coherently and which regions cancel. A large local overlap indicates strong local magnetoelastic interaction, whereas alternating phase or opposite domain sign reduces the net coherent integral.

The integrated projected overlaps are
\[
|I_{xy,i}^{2\mathrm{D}}|
=
5.66\times10^{-25},
\quad
3.55\times10^{-24},
\quad
2.53\times10^{-24}\,\mathrm{m^2}
\]
for $\mathrm{m_1}$, $\mathrm{m_2}$, and $\mathrm{m_3}$, respectively.
To quantify cancellation, we define
\begin{equation}
\eta_i
=
\frac{
\left|
\int_{A_{\mathrm{YIG}}}
\mathcal{I}_{xy,i}^{2\mathrm{D}}\,dA
\right|
}{
\int_{A_{\mathrm{YIG}}}
\left|
\mathcal{I}_{xy,i}^{2\mathrm{D}}
\right|dA
}.
\end{equation}
The values are
$\eta_1=1.49\times10^{-3}$,
$\eta_2=5.37\times10^{-3}$, and
$\eta_3=3.09\times10^{-3}$.
Thus, only about $0.15\%$, $0.54\%$, and $0.31\%$ of the absolute local overlap survives coherently for $\mathrm{m_1}$, $\mathrm{m_2}$, and $\mathrm{m_3}$, respectively; more than $99\%$ is cancelled by phase variation and stripe-domain sign reversal. This cancellation further explains the measured weak coupling. The idealized two-dimensional calculation reproduces the overall scale and mode dependence of the coupling, but not the detailed mode-by-mode ordering, possibly because it omits the $\epsilon_{yz}$ and $\epsilon_{zx}$ channels and assumes ideal, translationally invariant stripe domains~\cite{an2020coherent,an2023optimizing}. The overlap analysis nonetheless captures the central mechanism: shear strain couples locally to the stripe-domain magnon modes, while the measurable coherent coupling is strongly suppressed by phase variation and domain-sign cancellation.

We next discuss the consequence of this mechanism in a fully coupled YIG/GGG/YIG geometry, where phonons generated by stripe-domain magnons in one YIG layer can mediate a remote dynamic magnetization response across the GGG spacer. In this simulation, only the top YIG layer is directly excited by oscillating magnetic fields, while the bottom YIG layer responds through the elastic field generated by magnetoelastic coupling. Figure~\ref{fig:fig5}(a) shows the frequency dependence of $\delta m_x$ for both YIG layers, and we observe resonances in the driven top layer near $0.212$, $0.37$, $0.47$, and $0.53\,\mathrm{GHz}$. The excitation of magnetization dynamics in the bottom layer is also recognized by a weak but finite response in $\delta m_x$, demonstrating that the acoustic field launched from the top layer can remotely excite magnetization oscillation through the inverse magnetoelastic interaction.

It is important to highlight that the bottom-layer spectrum is not a scaled copy of the top-layer response. The large $\delta m_x$ in the bottom layer occurs at selected frequencies away from strong resonance peaks in the top layer. This reveals key aspects of magnetoelastic coupling in this configuration where the acoustic cavity must support phonon modes generated by magnetization oscillation. Indeed, the phonon spectrum plotted in Fig.~\ref{fig:fig5}(b) contains several sharp acoustic modes excited by the top layer, but only some of them couple efficiently to the bottom YIG.

The displacement profiles in Figs.~\ref{fig:fig5}(c)--\ref{fig:fig5}(e) help understand the origin of this. Figure~\ref{fig:fig5}(c) displays the spatial distribution of displacement at one of the resonance peak centers (0.212 GHz) where neighboring stripe domains are oppositely oriented in the ground state and therefore develop dynamic displacements of opposite signs when excited. At the resonance frequency, the displacement amplitude is resonantly enhanced within each domain, and the spatial distribution has odd symmetry along both vertical and horizontal directions. When integrated as a whole to generate propagating phonons, these contributions lead to efficient cancellation. On the other hand, away from the resonance center [Figs.~\ref{fig:fig5}(d) and (e)], the displacement becomes imbalanced across the domains, producing a finite net interfacial strain that propagates through the GGG channel. This explains why the largest remote response does not necessarily coincide with the largest top-layer magnon amplitude. It also identifies the condition for efficient propagating-phonon generation: only magnon modes with spatially asymmetric dynamic displacement produce a net interfacial strain that does not cancel upon integration, whereas symmetric profiles do not launch a propagating phonon.

The phonon-mediated coupling between the two YIg layers in our geometry remains weak. From Fig.~\ref{fig:fig5}(a), the bottom-layer dynamic magnetization amplitude is typically $10^{-4}$--$10^{-3}$ of the directly driven top-layer amplitude, corresponding to an intensity-like transfer ratio of approximately $10^{-8}$--$10^{-6}$ in the linear regime. This is consistent with the weak-coupling picture established experimentally: the phonons are long-lived, but the net drive transmitted across the spacer is reduced by domain-sign cancellation, imperfect interfacial strain balance, and mode-selective magnetoelastic coupling. Experimental realization of efficient spin-wave excitation mediated by long-lived phonons requires further work to enhance the magnetoelastic coupling strength and to optimize the coupling efficiency and phonon propagation characteristics in this geometry.

In conclusion, this work studies dynamical magnetoelastic coupling in a nonuniform magnetic ground state in YIG/GGG heterostructures. We observed phonon modes coupled to magnon modes in the stripe domains and quantified the coupling rates by a coupled harmonic-oscillator model. Fully coupled finite-element and micromagnetic simulations resolved the spatial profiles of the coupled magnon and shear-strain fields, identifying the overlap structure responsible for the weak, cancellation-limited coupling. We simulated the phonon-mediated dynamical coupling of two distant magnetic layers in the YIG/GGG/YIG structure and found that efficient generation of propagating phonons requires spatially asymmetric magnon modes. These results show that magnetoelastic coupling in domain-textured magnetic films can be controlled not only through macroscopic material parameters such as magnetic anisotropy, damping, and coupling strength, but also by engineering micromagnetic properties including the spatial distribution of static and dynamic textures. 

\begin{acknowledgments}
We acknowledge support by EPSRC under EP/X015661/1. D.P. is supported by the EPSRC and SFI Centre for Doctoral Training in Advanced Characterisation of Materials Grant Ref: EP/S023259/1. T.K. and E.S. acknowledge support from JST CREST (JPMJCR20C1 and JPMJCR20T2), Grant-in-Aid for Scientific Research (Grants No. JP19H05600, JP24K01326, and JP26K22729), and Grant-in-Aid for Transformative Research Areas (Grant No. JP22H05114) from JSPS KAKENHI, MEXT Initiative to Establish Next-generation Novel Integrated Circuits Centers (X-NICS) (Grant No. JPJ011438), Japan, and the Institute for AI and Beyond of the University of Tokyo. We thank Kei Yamamoto for valuable discussions.
\end{acknowledgments}

\bibliographystyle{apsrev4-2}
\bibliography{references}

\end{document}